%% file: main.tex
\documentclass[conference]{IEEEtran}
\IEEEoverridecommandlockouts
\usepackage{cite}
\usepackage{amsmath,amssymb,amsfonts}
\usepackage{algorithmic}
\usepackage{textcomp}
\usepackage{xcolor}
\usepackage{hyperref}
\usepackage{booktabs}
\usepackage{graphicx,txfonts}

\newcommand{\heart}{\ensuremath\varheartsuit}

\def\BibTeX{{\rm B\kern-.05em{\sc i\kern-.025em b}\kern-.08em
    T\kern-.1667em\lower.7ex\hbox{E}\kern-.125emX}}
\begin{document}

% \title{Exploring the \#WhyIDidntReport Movement to Understand the Silence of Sexual Violence Victims}
\title{Understanding the Silence of Sexual Harassment Victims Through the \#WhyIDidntReport Movement}

\author{
\IEEEauthorblockN{Abigail Garrett}
% \IEEEauthorblockA{\textit{Computer and Information Science} \\
\IEEEauthorblockA{\textit{Mathematics} \\
\textit{University of Mississippi}\\
agarrett@go.olemiss.edu}
\and
\IEEEauthorblockN{Naeemul Hassan}
\IEEEauthorblockA{\textit{Computer and Information Science} \\
\textit{University of Mississippi}\\
nhassan@olemiss.edu}
}

\maketitle

\begin{abstract}
Sexual violence is a serious problem across the globe. A lot of victims, particularly women, go through this experience. Unfortunately, not all of these violent incidents come to public. A large portion of victims don't disclose their experience. On the September of 2018, people started revealing in Twitter why they didn't report a sexual violence experience using a hashtag \#WhyIDidntReport. We collect about $40K$ such tweets and conduct a large-scale supervised analysis of why victims don't report. Our study finds the extent to which people shared their reasons as well as categorizes the reasons into finer reasons. We also analyze user engaged with the victims and compare our findings with existing literature.
\end{abstract}

% \begin{IEEEkeywords}
% Social Movement, Sexual Violence, Supervised Learning
% \end{IEEEkeywords}

\input{introduction.tex}
\input{literature_review.tex}

\input{method.tex}

\input{findings.tex}
\input{limitation.tex}

\bibliography{references}
\bibliographystyle{acm}

\end{document}

%% file: introduction.tex
\section{Introduction}
\label{sec:introduction}
Sexual violence is a serious offence that causes several negative impacts on the victims' physical, psychological, and social health~\cite{fitzgerald1997antecedents, loy1984extent}. Unfortunately, the prevalence of sexual violence is very high across the globe~\cite{gelfand1995structure}, and women are the principle victims of sexual violence. According to UN Women and World health Organization (WHO), $35\%$ (almost one in every three) of women worldwide have experienced either physical and/or sexual intimate partner violence or sexual violence by a non-partner at some point in their lives~\cite{unwomen}.

The U.S. Equal Employment Opportunity Commission~\cite{feldblum2016select} receives $12,000$ allegations of sex-based harassment each year. However, this number is believed to be just the tip of the iceberg. According to the Commission~\cite{feldblum2016select}, three of four sexual violence victims do not report to the authority.

Researchers~\cite{collegeSurvey, military, nationalcollege} and pertinent agencies (e.g., National Survey of Crime Victimization [NSCV]) have been trying to understand why victims do not report. Some of the common reasons which have been identified are- lack of information of where and how to report, fear of consequences, denial or minimization, etc. However, most of these studies are small-scale in nature and are focused to specific population. In September 2018, in light of the \textit{Brett Kavanaugh} hearing (\url{https://en.wikipedia.org/wiki/Brett\_Kavanaugh}), people started disclosing in Twitter why they didn't report their sexual violence experience using a hashtag \#WhyIDidntReport. It soon became viral and transformed into a movement. Victims from all over the United States participated in this movement. Their shared tweets gave an unprecedented opportunity of examining victims' silence in a larger scale with a broad range of population. In this paper, we analyze about $40,000$ tweets of this movement to conduct a large-scale analysis of victims' silence- reasons behind not reporting. First, we identify the most common reasons from existing literature. Then, we manually annotated a $10K$ sample of the data and use the annotated data to train supervised machine learning models. Finally, we use the models to detect and categorize the reasons from the full dataset. We find that Twitter users generally engaged more with the victims who broke their silence. We also observe that the distribution of reasons found by our study is aligned with some existing works. To the best of our knowledge, this is the first study that takes a social sensing approach to study the silence of the sexual violence victims. The observations from this study can be useful for sexual violence researchers, crime petrol entities, law and regulation authorities for developing victim friendly policies.

% Sexuhearal assault is a huge problem that greatly affects the victims throughout their lives.  \cite{consquence} studied whether childhood assault was related to PTSD or problem drinking later in life.  The study utilized a mail survey and concluded that childhood assault can result in PTSD about 30\% of the time and problem drinking 20\%.  Sexual assault is not an isolated event that victims experience and they move on from.  It can dictate how they perceive and interact with the world around them for the rest of their lives.  \cite{disclosing} discovered that when victims receive supportive reactions when they disclose their experience, they feel more control over their own recovery.  The mail survey indicated that most victims told at least one person and having support resulted in the use of more positive coping mechanisms.
\par
% The studies were done on sexual assault reveal a negative impact on victims lives.  Victims have methods available to deal with the trauma.  Hot lines and various forms of therapy can help a victim learn to cope with what happened to them.  However, having such a traumatizing experience affects many victims for the rest of their lives.  The best way for society to protect victims is to do everything to prevent future victims.  Understanding how prevalent the issue is and making sure victims feel as safe as possible disclosing their experiences are both positive first steps in creating a better culture surrounding assault victims.

% Look at other papers that work on understanding the reasons of silence. Their motivation can be used here.

% 1. to the best of our knowledge fist study on self report analysis
% 2. labeled data 
% 3. insights. copy from metoo
% 4. other studies are small. our is large scale

% conect with metoo

%% file: literature_review.tex
\section{Literature Review}
\label{sec:literature_review}
Related research that has been conducted on why victims do not report sexual violence generally takes a more focused approach. Some studies look at specific subgroups of victims~\cite{nationalcollege, military} or reporting~\cite{collegeSurvey} and most of them primarily use survey as the way for data collection. For instance, authors in ~\cite{nationalcollege, collegeSurvey} examined why college women don't report their assault by surveying college women nationally or by surveying women on a specific college campus.
% The National Survey focused on college women reporting to people other than authorities by using a survey with 4446 responses which was the largest sample size discovered from a survey research project \cite{nationalcollege}.  The University focused survey was interested in why women do not report to universities specifically and had a sample size of 220 \cite{collegeSurvey}.
Both studies reached similar conclusion as to why, in these cases women, rarely report.  They reference that the women were afraid, ashamed, did not think it was bad enough, or were drugged or drinking. The authors in~\cite{nationalcollege} also suggests that a relationship to the offender and personal attributes like race, age, and gender can make a woman less likely to report.  The college women study~~\cite{collegeSurvey} mentions that women did not know they could report, or did not report to the university because it was not university-related. The authors in~\cite{military} focus specifically on service women and also uses a survey which was mailed to participants.  The study drew on about $2,800$ participants which included those that chose not to respond. 
% It presents another example of a larger survey that again narrows the focus to a specific group of people.  However, \cite{interview} present a slightly different approach.  The study is slightly different than the topic of this thesis as it focuses on why victims may not label their experience as assault.  However, in the end, the result focuses on how this can affect reporting.  It brings a new approach because \cite{interview} uses a mixed form of data retrieval.  The study gathers data using interviews, focus groups, and observation of subjects.  It also looks at undergraduates, but the difference in data gathering presents a new approach to the subject.
Yet another approach, M{\'e}nard \cite{ageBook} studies the question using data gathered from $48$ rape crisis centers from Pennsylvania. It focuses more on cultural aspects of not reporting. This study takes a broader approach rather than focusing on certain population group. Also, unlike previous studies, we do a large-scale analysis of about $10K$ self-reported reasons instead of conducting survey of focus group study.

%% file: method.tex
\section{Methodology}
\label{sec:method}

\subsection{Data Preparation}
In this section, we describe our data collection strategy, the annotation process, and observations from the annotated data.

\subsubsection{Collection}
\label{sec:data_collection}
\begin{figure}[ht!]
    \centering
    \includegraphics[width=0.9\linewidth]{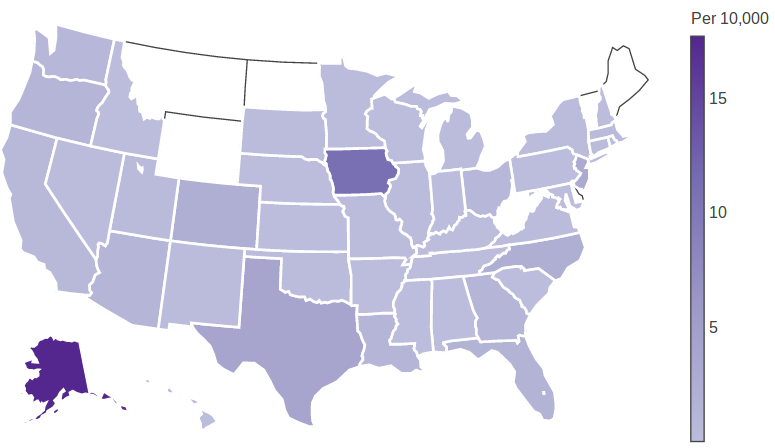}
    \caption{Number of tweets per 1000 people from the U.S. states}
    \label{fig:usmap}
\end{figure}

We collected about $40,000$ tweets containing the \emph{\#WhyIDidntReport} hashtag from all the U.S. cities that had a population of more than $100K$ using the Twitter Advanced Search Feature. These tweets were collected within the September $22$ -- October $2$, $2018$ period. This is the period when \emph{Brett Kavanaugh hearing}~\footnote{\url{https://en.wikipedia.org/wiki/Brett_Kavanaugh_Supreme_Court_nomination\#Hearing}} was going on and the \#WhyIDidntReport movement was in its peak. The tweets contained a variety of
subject matters that ranged from commentary on the Kavanaugh hearings to personal accounts of
assault to reasons why victims did not report the assault.
% We have made a $25\%$ random sample of the whole dataset publicly available for other researchers ~\footnote{Link is not provided due to anonymous submission policy.}.
We removed the duplicate tweets from the dataset by matching the permanent URLs of the tweets. This reduced the collection size to $37,526$ tweets that were tweeted by $24,194$ unique users. Figure~\ref{fig:usmap} shows a choropleth indicating the number of tweets per $10,000$ people in different U.S. states. The figure shows that there was participation in the movement from almost all over U.S. In total, the dataset covers $211$ cities from $43$ states. The top three states with highest tweet density are- \emph{Alaska}, \emph{Iowa}, and \emph{Texas}.

\subsubsection{Annotation Scheme}
Our objective was to identify the tweets that disclose a reason for remaining silent after a sexual violence. So, it was necessary to carefully prepare a tweet annotation guideline that conforms with existing literature. To prepare an annotation scheme, first we conducted a pilot study with $1,000$ random tweets. We observe that a large portion of the tweets did not actually share reasons. So, to increase the probability of finding tweets with reasons, we filtered the data with a set of keywords (\emph{`because', `I was', `I thought', `I felt', `I didn't'}) which frequently appeared in reason mentioning tweets according to our pilot study. This step reduced the dataset size to $12,899$ tweets.

Standing on existing literature that studied victims' reasons for not disclosing sexual violence experience~\cite{nationalcollege, collegeSurvey, military, psychtoday}, we prepare a list of eight reasons-

\begin{enumerate}
\item
\textbf{Shame}: The victims feel that it was their own fault and they were too ashamed to tell anyone.
\item
\textbf{Denial} or Minimization: The victims were trying to convince themselves that what happened really was not a big deal, did not happen, or was not wrong.
\item
\textbf{Fear} of Consequences: This category is rather broad because it can be a fear of losing one’s job, what people will think, the actual legal aspects of reporting, physical harm from the assailant or others, or anything else that could cause the fear which prevents the victim from reporting.
\item
\textbf{Hopeless}ness or Helplessness: Victims feel as though there is no point in reporting because they have seen how people treat those who do report and have seen the lack of action when an assault is reported, especially due to others' disbelief.
\item
\textbf{Drug}/Disassociation: Substance or psychological effects prevent the victim from having a clear memory of the event.
\item
\textbf{L}ack \textbf{o}f \textbf{I}nformation: The victim didn't have information of where or who to report, how to report. Or, the victim didn't know that the assault can be reported.
\item
\textbf{Protect}ing Assailant: The victims do not want the assailant to go to jail or have their life ruined.
\item
Young \textbf{Age}: The victims didn't report because they were young.
\end{enumerate}

A undergraduate female student manually analyzed the tweets and labeled if the tweet mentions a reason. After that, she categorized the reason to one of these eight cases. A tweet sometimes provide multiple reasons. In that case, she categorized the tweet into multiple reasons.

\textbf{Example: } ``\textit{I was 15 and was told if I told anyone or didn't go through with it I didn't love him and he would leave me. I didn't want to get in trouble and I didnt think it counted as rape. Years down the it still effects my life.}'' For example, this tweet mentions multiple reasons-- $Fear$ (\emph{I didn't want to get in trouble}) and $Denial$ (\emph{I didnt think it counted as rape}).

\subsubsection{Annotated Data}
In total, we identified $6,860$ tweets that contained at least one reason and $3,935$ tweets that didn't mention any reason for not reporting earlier. Table~\ref{tab:reason_distribution} shows the number and percentage of tweets in each reason category. The most common reason is feeling hopeless or helpless- about $30.877\%$. The least common reason is to protect the image of the assailant, $2.545\%$. There are $4,286$ tweets that mentioned only one reason, $1,789$ tweets mentioned two, and $570$ tweets mentioned three or more reasons.

\subsection{Problem Formulation}
We model the reason understanding problem as a two step process- \textbf{i)} detecting tweets that contain a reason for not disclosing sexual harassment experience \textbf{ii)} categorizing the detected tweets into appropriate reasons.

\subsubsection{Reason Detection}
\label{sec:detection}
We define the reason providing tweet detection task as a supervised binary classification problem where the set of classes, $\mathcal{RD} =  \{Reason, no\_Reason\}$. Formally, given a tweet $t$, the goal is to map $t$ to $c$ where $c \in \mathcal{RD}$.

\subsubsection{Reason Categorization}
\label{sec:categorization}
In a tweet, a victim may describe multiple reasons for being silent, in other words, for not disclosing sexual violence experiences. So, we define the reason categorization task as a multi-label supervised classification problem where the set of classes, $\mathcal{RC} =  \{Shame$, $Denial$, $Fear$, $Hopeless$, $Drug$, $LoI$, $Protect$, $Age\}$. Formally, given a tweet $t$ that has been mapped to $Reason$, the goal of this task is to map $t$ to $C$ where $C \subseteq \mathcal{RC}$.

\subsection{Model Development}
Before applying machine learning models, we preprocessed the tweets following a series of steps. First, we converted the tweets to all lowercase. Then, a Python package named tweet-preprocessor~\footnote{https://pypi.org/project/tweet-preprocessor/} was used to remove hashtags, URLs, emojis, and smileys from the tweets. After that, these cleaned tweets were transformed to vectors using Bag-of-Words (BOW) model and TF-IDF (with sublinear tf). We used unigrams and bigrams as features (ngrams with n = 1, 2). There were $117,274$ such features. The ngrams that were present in only one tweet (very rare) or present in more than $25\%$ of all tweets (very common) were discarded from feature set. This step reduced the number of features to $23,375$. From these, we selected the most frequent $5,000$ ngram features and used these in all the experiments.

We experimented with multiple supervised learning algorithms. Specifically, we used Support Vector Machine (SVM) with linear kernel, Random Forest, Naive Bayes, and Gradient Boosting. In most cases, the SVM outperformed other models. So, in this paper, we only reported performance of SVM. In all experiments, we split the data into train and test sets with a 80:20 ratio and use the test set for evaluation. In the \textbf{Reason Detection} task, the SVM model achieves an average (weighted) precision, recall, and F1-measure of $90\%$. Table~\ref{tab:rd_performance} shows detailed performance for each class. The SVM based reason detector can detect reason mentioning tweets with a $92\%$ precision and $93\%$ recall.

\vspace{-3mm}
\begin{table}[h!]
\centering
\caption{Performance of Reason Detector}
\begin{tabular}{l|llr}
\toprule
             & Precision & Recall    & F1-measure\\
\midrule
Reason        & 0.92      & 0.93      & 0.92  \\
no\_Reason    & 0.87      & 0.84      & 0.86  \\
\midrule
Avg (weighted)& 0.90      & 0.90      & 0.90  \\
\bottomrule
\end{tabular}
\label{tab:rd_performance}
\end{table}

To further understand the quality of the reason detector we conducted a ranking evaluation using the class probabilities of the classified tweets. We order the test set tweets with respect to their SVM assigned \textit{Reason} probability and measured the quality of the top-$k$ tweets by two commonly used metrices- Precision-at-k (P@k), and Average Precision-at-k (AvgP@k). Table~\ref{tab:rd_ranking} shows these measures for various values of $k$. In general, the detection model achieves excellent performance in ranking. The precision of reason detection among the top $1000$ tweets is $97.3\%$. 

\begin{table}[!ht]
  \centering
    \caption{Ranking evaluation of Reason Detection}
    \resizebox{\linewidth}{!}{%
    \begin{tabular}{@{}l|rrrrrrrr@{}}
    \toprule
    k      & 25 & 50 & 100 & 300   & 500   & 1000   &   1500    &   2000\\
    \midrule
    P@k    & 1.0  & 0.98  & 0.99   &  0.997 &   0.996   &   0.973   &   0.886   &   0.697 \\
    \midrule
    AvgP@k & 1.0  & 0.997  & 0.992   & 0.994 & 0.995 & 0.991 & 0.980  &   0.973 \\
    \bottomrule
    \end{tabular}%
    }
    \label{tab:rd_ranking}
\end{table}

We model the \textbf{Reason Categorization} as a multi-label classification problem. For each $c \in \mathcal{RC}$, we develop a one-vs-rest SVM classifier. Table~\ref{tab:rc_performance} shows performance of each category in terms of precision, recall, F1-measure, and accuracy. The F1-measures for 
\emph{Shame}, \emph{Fear}, and \emph{Hopeless} are above $75\%$. However, on the rest of the categories, the classifiers didn't perform well. Particularly, for the \emph{Age} category, the F1-measure is worse than a random guess. The could be due to the lack of samples from these classes. We plan to continue annotating more tweets that can potentially increase the sample size and thereby lead to a better result.

\vspace{-3mm}
\begin{table}[h!]
\centering
\caption{Performance of Reason Classifiers}
\begin{tabular}{l|lllr}
\toprule
Reason  &   Precision   &   Recall  &   F1-measure  & Accuracy\\
\midrule
Shame & 0.85  & 0.728 & 0.789 & 0.896 \\
Denial & 0.77  & 0.331 & 0.551 & 0.921 \\
Fear  & 0.819 & 0.713 & 0.766 & 0.877 \\
Hopeless & 0.794 & 0.76  & 0.777 & 0.813 \\
Drug & 0.75  & 0.296 & 0.523 & 0.958 \\
LoI   & 0.711 & 0.408 & 0.56  & 0.907 \\
Protect & 0.76  & 0.422 & 0.591 & 0.977 \\
Age   & 0.662 & 0.293 & 0.477 & 0.908 \\
\bottomrule
\end{tabular}
\label{tab:rc_performance}
\end{table}

If a tweet is labeled positive by multiple one-vs-rest models, we assign the corresponding classes to that tweet. We use \textit{Hamming Loss}~\cite{dembczynski2010regret} (HL) to measure the quality of the multi-label assignments. HL is the fraction of labels that are incorrectly predicted. The equation for measuring HL is as follows-

$$HL = \frac{1}{N\times L}\sum_{i=1}^{|N|}\sum_{j=1}^{|L|}\oplus~(y_{i,j}, z_{i,j})$$

where $N, L, y, z$ are the number of tweets, number of labels (categories), true label, and predicted label, respectively and $\oplus$ denotes an \textbf{xor} operation. HL is lower (close to zero) the better. Our one-vs-rest classification models achieved a very low HL score of $0.093$.

%% file: findings.tex
\section{Findings}
\label{sec:findiings}
Using the above described classifiers, we conduct an exploratory analysis on the \#WhyIDidntReport movement data. The data collection method for the \#WhyIDidntReport has been explained in Section~\ref{sec:data_collection}. The whole corpus had $37K$ tweets of which about $11K$ tweets were annotated for training and evaluating the classification models and the remaining $26K$ unlabeled tweets are analyzed by the trained classifiers. Our goal is to understand to what extent sexual violence victims were silent and what were the reasons.

\textbf{Extent: }Our reason detector identifies $13,612$ tweets that mention at least one reason for not reporting a sexual violence earlier. To conduct a robust analysis, we identified the reason mentioning tweets that have \emph{Reason} probability greater than or equal to $0.8$. The detector model has high confidence on these tweets. This threshold is set empirically, after attempting with multiple smaller and larger values. There are $9,866$ self-reports which surpass this threshold. We continue our analysis with these tweets only. There are $5,882$, $2,377$, and $524$ tweets that contain one, two, and three or more reasons, respectively.

\vspace{-2mm}
\begin{table}[ht!]
\centering
\caption{Frequency and Percentage of Reasons in Annotated and SVM\_labeled Data}
\resizebox{\linewidth}{!}{%
\begin{tabular}{@{}l|ll|lr@{}}
\toprule
Reason   & \# in Annotated & \# in SVM\_labeled & \% in Annotated & \% in SVM\_labeled \\ \midrule
Shame    & 1811      & 2450       & 18.740       & 19.974        \\
Denial   & 680       & 647        & 7.036        & 5.275         \\
Fear     & 1988      & 2650       & 20.571       & 21.604        \\
Hopeless & 2984      & 4329       & 30.877       & 35.293        \\
Memory   & 400       & 317        & 4.139        & 2.584         \\
LoI      & 829       & 951        & 8.578        & 7.753         \\
Protect  & 246       & 272        & 2.546        & 2.218         \\
Age      & 726       & 650        & 7.512        & 5.299         \\
\bottomrule
\end{tabular}%
}
\label{tab:reason_distribution}
\end{table}

We apply the reason categorization model on the $9,866$ high-confidence reason mentioning tweets. Table~\ref{tab:reason_distribution} shows the distribution of the reason categories. Overall, the distribution in the annotated data and the SVM\_labeled data is are similar. \emph{Hopelessness and Helplessness} is the most common and \emph{Protecting Assailant} is the least common reason for not reporting sexual violence incident.

\textbf{User Engagement: } We study how the Twitter users engage with the victims when they break their silence and disclose the sexually violent experience. Specifically, we measure how many \textit{Like} (\heart) a reason mentioning tweet receives. Table~\ref{tab:user_engagement} shows the average and median number of \textit{Likes} a tweet got depending on it mentioned a reason or not. We observe that a reason mentioning tweet receives \textit{Like} about two times more than an average tweet. On average, a reason mentioning tweet got $7.08$ \textit{Likes} and other tweet got $3.755$ \textit{Likes} in our annotated data (in the SVM\_labeled data, the numbers are $6.603$ and $3.698$, respectively). Note, we discarded the highly popular tweets ($Likes \ge 100$) to avoid outlier from this analysis. In summary, the Twitter users engaged more with the reason mentioning tweets than with other tweets. 

\vspace{-2mm}
\begin{table}[h!]
\centering
\caption{User Engagement with \textit{Reason} and \textit{no\_Reason} Tweets}
\begin{tabular}{@{}l|cc|cc@{}}
\toprule
        & \multicolumn{2}{c}{Annotated} & \multicolumn{2}{c}{SVM\_labeled} \\ \midrule
        & \textit{Reason}      & no\_Reason      & Reason        & \textit{no\_Reason}       \\
        \midrule
Average & 7.083       & 3.755           & 6.603         & 3.569            \\
Median  & 2           & 1               & 2             & 1                \\ \bottomrule
\end{tabular}
\label{tab:user_engagement}
\end{table}

\textbf{Comparison with Other Studies: } There are existing works that tried to understand the reasons for not reporting. Unlike this work, they collected data through survey or interview. For instance, Truman and Langton~\cite{truman2015criminal} found that fear and hopelessness are common reasons for not reporting sexual violence. Our study also finds these as the two most common reasons~\ref{tab:reason_distribution}. Another study~\cite{nationalcollege} also finds \emph{Denial}, \emph{Hopeless} and \emph{Fear} to be the most common three reasons. However, we didn't find \emph{Denial} as a common reason. Spencer et al.~\cite{collegeSurvey} found that the most common reasons are \emph{LoI} and \emph{Denial} which are different from our most common reasons. The difference could stem from the fact that they studied a specific population- college or university students.

%% file: limitation.tex
\section{Conclusion, Limitations and Future Work}
\label{sec:limitations}
In summary, this paper studies why the victims remain silent after going through sexual violence. The findings can be useful for sexual violence researchers, crime petrol entities, law and regulation authorities for developing victim friendly policies. While this study reveals some insights from the \#WhyIDidntReport movement data, there are some areas for possible improvement. First, the data was carefully annotated by one female student. This can introduce personal bias in the data. One possible solution is to annotate the data using more people. Second, the annotated data didn't have enough samples for some of the categories. Third, a tweet has $240$ character length limit. Tweeters often bypass this by using multiple tweets to make a single long post. As we treated each tweet as a single post, these kind of long posts weren't considered as a single unit which can potentially lead to inaccurate results. In future, we plan to resolve these limitations by deploying more annotators and designing robust tweet processing techniques.
% Also, we plan to contextualize the reasons with respect to victims' features such as gender, culture, age, etc.